# Interaction of OAM light with Rydberg excitons: Modifying dipole selection rules

Annika Melissa Konzelmann,[1] Sjard Ole Krüger,[2] and Harald Giessen[1*]

[1] *4th Physics Institute and Research Center SCoPE, University of Stuttgart, Pfaffenwaldring 57, D-70569 Stuttgart, Germany*
[2] *Institut für Physik, Universität Rostock, Albert-Einstein-Straße 23, D-18059 Rostock, Germany*
[*] *h.giessen@pi4.uni-stuttgart.de https://www.pi4.uni-stuttgart.de*

**Abstract:** Orbital angular momentum (OAM) light possesses in addition to its usual helicity ($s = \pm\hbar$, depending on its circular polarization) an orbital angular momentum $l$. This means that in principle one can transfer more than a single quantum of ħ during an optical transition from light to a quantum system. However, quantum objects are usually so small (typically in the nm range) that they only locally probe the dipolar character of the local electric field. In order to sense the complete macroscopic electric field, we utilize Rydberg excitons in the semiconductor cuprite ($Cu_2O$), which are single quantum objects of up to µm size. Their interaction with focused OAM light, allows for matching the focal spot size and the wavefunction diameter. Here, the common dipole selection rules ($\Delta j = \pm 1$) should be broken, and transitions of higher $\Delta j$ with higher order OAM states should become more probable. Based on group theory, we analyze in detail the optical selection rules governing this process. Then we are able to predict what kind of new exciton transitions (quantum number $n$ and $l^{exc}$) one would expect in absorption spectroscopy on $Cu_2O$ using different kinds of OAM light.

## I. INTRODUCTION

The giant size of Rydberg atoms leads to huge interaction effects from which one gains insight into atomic physics on the single quantum level.[1,2] In the solid state there exists an analog, called Rydberg exciton, which is an exciton state with large principal quantum number $n$. This exciton should be capable of sensing elementary excitations in its surrounding on a quantum level.[3] An exciton is an excited state of the crystal, in which an electron and a hole form a quasiparticle bound by Coulomb interaction. In covalent crystals, such as cuprous oxide ($Cu_2O$), excitons are delocalized, the electron-hole pair is loosely bound, and the orbits are large with macroscopic dimensions of ~1µm. The so-called Wannier excitons appear in the low energy spectrum of the crystal as sharp absorption peaks below the bandgap.[4] Excitonic effects are decisive for the optical properties of semiconductors,[5] among which cuprous oxide is unique in crystal quality.[6] Rydberg excitons are well suited for the investigation of interaction effects. Huge polarizabilities are expected, leading to enormously strong dipolar interactions.[7] Furthermore, in contrast to Rydberg atoms, highly excited excitons with µm-size extensions are of interest because they can be placed and moved in a crystal with high precision using macroscopic energy potential landscapes.[8]

The energies of the optically excited excitons can be determined directly by one-photon absorption studies, in which the photon energy of a single-frequency laser with a narrow spectral linewidth of a few neV is continuously tuned. The exciton energy series can be calculated according to: $E_n = E_g - E_B$, with the band gap energy $E_g = 2.17\ eV$, the exciton binding energy $E_B = Ry^*/(n - \delta_{n,l})^2$, the quantum defect $\delta_{n,l}$ induced by the non-parabolic hole dispersions,[9] and the modified Rydberg constant[10] $Ry^* = Ry\, m^*/(\varepsilon^2 m)$. The crystal environment is taken into account through the permittivity ε, which is isotropic for cubic ($O_h$) symmetry, and the effective electron and hole masses $m_e$ and $m_h$ are incorporated via $m^* = m_e m_h/(m_e + m_h)$.

We are interested in excitons which are formed between the highest valence and lowest conduction bands in cuprous oxide (see FIG. 1 (a)). These bands are formed by copper 3d and 4s orbitals, respectively. Both bands have the same parity, thus the transition dipole moment for band-to-band transitions vanishes. When forming excitons, only so-called second-class transitions are possible. This means that beyond the valence and conduction band symmetry, also the symmetry of the exciton envelope wavefunction has to be taken into account. This means that for our case of s- and d-type valence and conduction bands ($\Delta l = 2$), only P-excitons with envelope angular momentum $l^{exc} = 1$ can be excited. In this case, the photon carries one ħ of angular momentum (RCP or LCP light), and the p-envelope of the exciton carries the second ħ of angular momentum that is necessary to make the transition with $\Delta l = 2$ from the 3d valence to the 4s conduction band. Hence, in $Cu_2O$, electric dipole transitions cannot lead to excitons with s-type envelope. However, the transition dipole moment of the second-class transition (p-type envelope) is typically one order of magnitude smaller than for s-type excitons in semiconductors with s- and p-type bands (e.g., GaAs).

All aforementioned considerations are usually done for plane waves with zero *orbital* angular momentum. However, light beams with an azimuthal phase dependence of $e^{-il\varphi}$ carry an additional orbital angular momentum $l\hbar$, independent of their polarization state.[11–15] For any given $l$ the beam has $l$ intertwined helical phase fronts, for which the Poynting vector has an azimuthal component, meaning it is no longer parallel to the beam axis. That component produces an orbital angular momentum parallel to the beam axis, which is associated with regions of high intensity. This comes together with a phase singularity on the beam axis with zero optical intensity and no linear or angular momentum, which persists no matter how tightly the beam is focused. The most common forms of helically phased beams are the Laguerre-Gaussian laser modes. Such donut modes form a complete basis set for paraxial light beams and have circular symmetry. Orbital angular momentum light can be straight-forwardly produced using computer generated holograms[15] or spiral phaseplates.[16] In addition, OAM light carries a strong field gradient in its center while there is no E-field intensity (dark penumbra in the beam center). This unique feature makes the interaction of OAM light with *atomic* matter different from plane waves.[17] In particular, during the interaction of OAM light and atoms one can not only transfer the physical properties of OAM light (spin $s$ and angular momentum $l$) to internal and external degrees of freedom,[18] but also can OAM beams trap particles due to their field gradient.[19]

Most importantly, with the combination of OAM light and Rydberg excitons in cuprite (macroscopic quantum objects), we bridge the size gap between light and matter.[20] Different from atomic physics, the size of the Rydberg excitons, which are single quantum objects, is comparable to the size of a tightly focused OAM beam. In fact, the $n = 25$ Rydberg exciton has a Bohr radius of 920 nm, whereas the wavelength for its excitation is 571 nm, which can be focused down to a tight focus of about 500 nm. *This gives us for the first time the unique opportunity to investigate the interaction of OAM beams with single quantum objects with comparable sizes in the hundreds of nanometer range.* In atomic and ionic systems, the small sizes in the Ångström range probe the OAM beam only locally and thus experience a local field with mostly dipolar character.

When we regard crystals as macroscopic continua, the rotational symmetry is broken down to discrete groups.[21] However, in cuprous oxide this rotational symmetry is still quite high (point group $O_h$). The symmetry transformations of a Hamiltonian always form a group, which is directly related to the physical symmetry of the system to which the Hamiltonian applies. Such symmetry considerations can be used to extract information residing in the respective transition matrix elements regarding the selection rules associated with orbital angular momentum transfer. Optical transitions occur from the crystal ground state to the excitonic states and are driven by the light field operator. Exciting matter with OAM light allows for one more degree of freedom (spin *and* angular momentum quantum number) compared to usual dipolar light (spin quantum number only). Applied to Rydberg excitons, which can be in different angular momentum quantum states, we can engineer new selection rules with OAM light, thus, having a rich set of control parameters.[22]

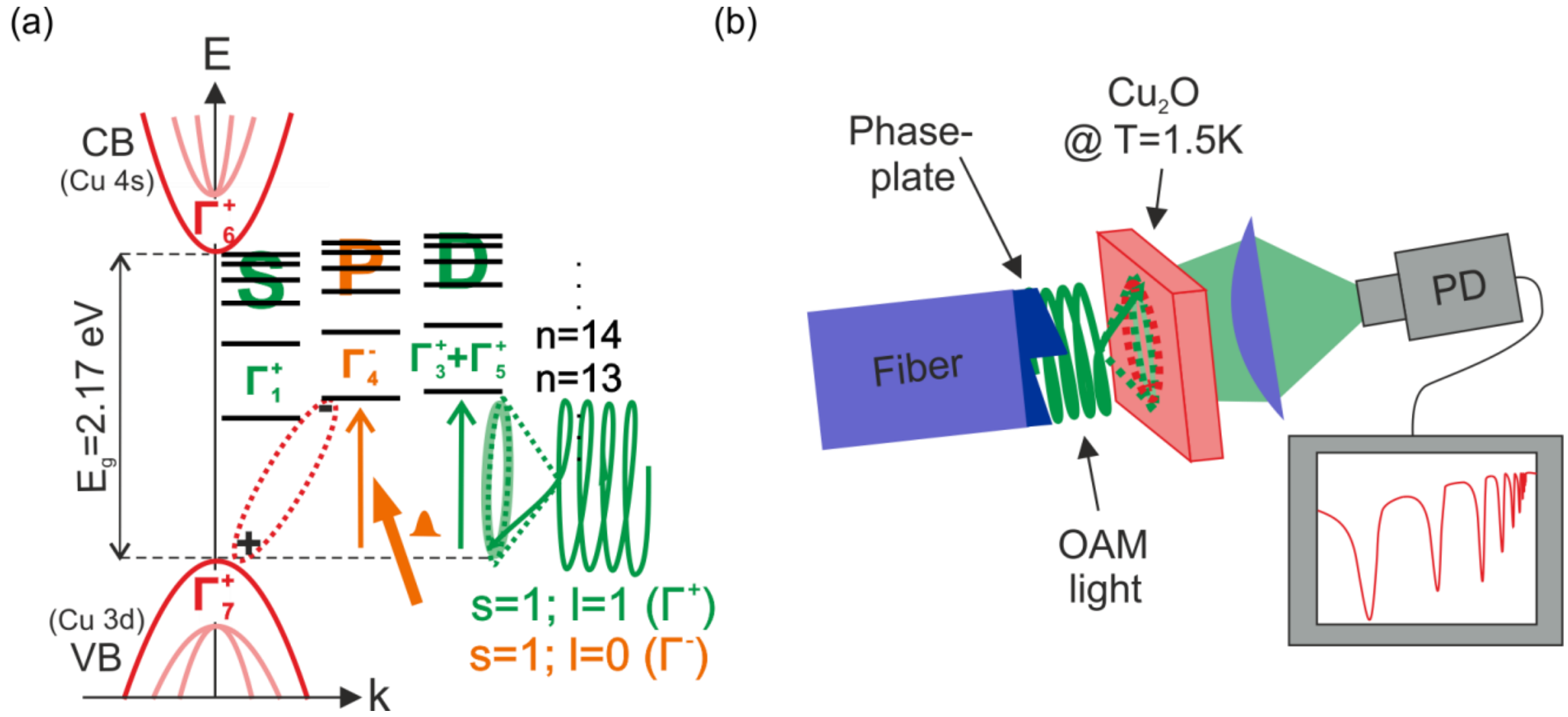

FIG. 1. (a) Formation of different exciton series between the highest valence (VB) and the lowest conduction (CB) bands in $Cu_2O$ depending on the light properties ($s = 1$; $l = 0,1$). Using OAM light, Rydberg excitons with different envelope functions ($S$, $P$, $D$, ...) can be excited by allowed transitions. (b) Experimental scheme: A spiral phaseplate imprinted on a fiber facet creates OAM light, which excites Rydberg excitons in cuprite at cryogenic temperatures. The signal is detected using a photodiode (PD). The Rydberg exciton series is visible in the absorption spectrum as broadened, asymmetric, Fano-shaped absorption lines.

## II. METHODS

Our aim is to predict whether an optical transition between the cuprite crystal ground state and exciton states with different amount of angular momentum $l^{exc}$ is allowed or forbidden when exciting with OAM light. For allowed transitions the overlap of crystal ground and excited state contains the symmetry of the exciting light (optical transition driving operator $\boldsymbol{A} \cdot \boldsymbol{p}$). The symmetries of the excitonic states are derived from the cuprous oxide band symmetries. As the excitation takes place inside the crystal, we use the tables of Ref. [23] in order to determine the OAM light symmetries within the symmetry group of cuprite.

Cuprous oxide belongs to symmetry group $O_h$, which contains all 24 proper and 24 improper rotations which transform a cube into itself. See Supplemental Material at [*URL*] for a detailed description of this point group.[24] Knowing how the symmetry group behaves under parity, it suffices to take only the symmetry classes with the proper rotations into account. The symmetry of an exciton is described by the product of the symmetries of the conduction band (electron), the valence band (hole), and the envelope function, determining the angular momentum quantum state. The former two symmetries are known from literature[25] and the latter one can be derived from the atomic orbitals of the hydrogen wave function due to the high symmetry of cuprite.[26-28] Orbital angular momentum light is described by a Laguerre-Gauss mode, which can be decomposed, in the longitudinal dipole approximation, into a plane wave part $\boldsymbol{A_0}$ times a phase factor $e^{il\varphi}$. While the symmetry of the plane wave part is trivial, in order to extract the symmetry of the phase factor, we decompose the latter one into basis functions, analyze how these basis functions change when undergoing $O_h$ symmetry operations and calculate the trace of the transformation matrix (characters). For the complete set of characters (one character per symmetry operation) the corresponding light field symmetry can be assigned according to the character tables of group $O_h$.[23]

## III. RESULTS AND DISCUSSION

### A. OAM light symmetry assignment via calculation of transformation matrix diagonal elements

In the following we are going to assign a symmetry to the optical transition driving operator involving OAM light modes with different amount of orbital angular momentum $l$. OAM light is described by a Laguerre-Gauss mode which is a Gaussian beam times an additional phase factor $e^{il\varphi}$. The complete light field operator $\boldsymbol{A}$ has the form:

$$\boldsymbol{A_{lp}}(r,\varphi,z) = \boldsymbol{A_0}\, e^{ikz} \frac{w_0}{w} \exp\left(\frac{-r^2}{w^2} + \frac{ikr^2}{2R} - i(2p + |l| + 1)\phi(z)\right) \left(\frac{\sqrt{2}r}{w}\right)^{|l|} L_p^{|l|}\left(\frac{2r^2}{w^2}\right) e^{il\varphi}, \quad (1)$$

with $\boldsymbol{A_0}$: light field amplitude, $k$: wave vector, $w(z)$: beam waist, $w_0 = w(z=0)$, $\phi(z)$: Gouy phase, $R(z)$: radius of curvature, $L_p^{|l|}$: generalized Laguerre polynomial.

We consider in our calculations only the case where the OAM light propagates in $z$-direction. The exciton is assumed to be located in the vortex center ($xy$-plane), so the overall symmetry is maintained and the r-dependence of the mode does not influence the transformational properties. However, as the field gradient in the OAM light beam becomes important for quadrupole transitions, we develop the light field in $z$. Then, in dipole approximation the Gaussian beam part simplifies to $\boldsymbol{A_0}\, e^{ikz} \approx \boldsymbol{A_0}\,(1 + ikz) \approx \boldsymbol{A_0}$. From the OAM part, the only relevant factor for symmetry considerations is $e^{il\varphi}$. Then, the optical transition driving operator becomes $\boldsymbol{A} \cdot \boldsymbol{p} = \boldsymbol{A_0} e^{il\varphi} \cdot \boldsymbol{p}$. The symmetries of $\boldsymbol{A_0}$ and $\boldsymbol{p}$ with respect to group $O_h$ are $\Gamma_1^+$ and $\Gamma_4^-$, respectively. The function $e^{il\varphi}$ is cylindrically symmetric and can be aligned along the three coordinate axes in the cubic cuprite crystal. Thus, we consider the six linearly independent basis functions $e^{+il\varphi_x}$, $e^{-il\varphi_x}$, $e^{+il\varphi_y}$, $e^{-il\varphi_y}$, $e^{+il\varphi_z}$, $e^{-il\varphi_z}$, into which $e^{il\varphi}$ can be transformed under the symmetry operations of point group $O_h$. For an orientation along different axes, the number of linearly independent basis functions would differ, thus the following analysis only holds strictly for the case in which the beam axis aligns with a coordinate axis. In order to determine the symmetry of function $e^{il\varphi}$, we analyze how the basis functions change when undergoing $O_h$ symmetry operations, and calculate the trace of the transformation matrix (characters). For the complete set of characters, the corresponding symmetry can be assigned according to the character tables of group $O_h$. The case $l = 0$ needs to be considered separately. As $e^{i(l=0)\varphi} = 1$, we cannot find six linear independent basis functions, contrary to the case $l > 0$. The symmetry for the dipolar light field ($l = 0$) is $\Gamma_1^+$.

A coordinate transformation of order $n$ leads to the original coordinate after performing the transformation $n$-times, i.e., three times for the eight three-fold symmetry axes $8C_3$ of group $O_h$: $x \to y \to z \to x$. It suffices to select one element per class, i.e., axis (111) in class $8C_3$, as all elements of the same class transform the same way. In contrast, it is important to perform the symmetry considerations for a complete basis, i.e., $x$, $y$, and $z$, in order to extract the character of a transformation (trace of transformation matrix). The complete transformations of the coordinates $x$, $y$, $z$ under symmetry operations of group $O_h$ are listed in Table S1 in the Supplemental Material.[24] Knowing how the coordinates behave under the different symmetry operations, one can write down the transformations of the basis functions of $e^{il\varphi}$ (see Table I).

**TABLE I. Transformations of functions $e^{\pm il\varphi_x}$, $e^{\pm il\varphi_y}$, and $e^{\pm il\varphi_z}$ under $O_h$ symmetry operations.**

| Operator $\hat{O}$ | $\hat{O}\lvert e^{\pm il\varphi_x}\rangle$ | $\hat{O}\lvert e^{\pm il\varphi_y}\rangle$ | $\hat{O}\lvert e^{\pm il\varphi_z}\rangle$ |
|---|---|---|---|
| E | $e^{\pm il\varphi_x}$ | $e^{\pm il\varphi_y}$ | $e^{\pm il\varphi_z}$ |
| $8C_3$ | $e^{\pm il\varphi_z}$ | $e^{\pm il\varphi_x}$ | $e^{\pm il\varphi_y}$ |
| $3C_2$ | $e^{\pm il(\varphi_x+\pi)}$ | $e^{\mp il(\varphi_y+\pi)}$ | $e^{\mp il\varphi_z}$ |
| $6C_4$ | $e^{\pm il(\varphi_x+\pi/2)}$ | $e^{\mp il(\varphi_z\pm\pi/2)}$ | $e^{\pm il(\varphi_y\mp\pi/2)}$ |
| $6C_2^1$ | $e^{\pm il(\varphi_y-\pi/2)}$ | $e^{\pm il(\varphi_x+\pi/2)}$ | $e^{\mp il(\varphi_z-\pi/2)}$ |

In order to calculate the transformation matrix diagonal elements, the transformed functions are decomposed into basis vectors according to $\hat{O}|f_i\rangle = \sum_k N_{i,k}|f_k\rangle$. The characters of the OAM light for each symmetry class are then given by the sum of the results of all six basis vectors, i.e., the trace $\chi(\hat{O}) = Tr(N_{i,k}) = \sum_k N_{k,k}$. The single transformation matrix diagonal elements for one representative class element of all symmetry classes of group $O_h$ as well as the resulting character set for OAM light ($e^{il\varphi}$) are listed in Table S2 in the Supplemental Material.[24] Multiplication of the characters of $e^{il\varphi}$ with the ones of $\boldsymbol{A_0}$ yield the character set of the complete OAM light field operator $\boldsymbol{A} = \boldsymbol{A_0} e^{il\varphi}$. The amplitude $\boldsymbol{A_0}$ is described by symmetry $\Gamma_1^+$, which acts like a 1 in multiplication (identity). Further multiplication with the characters of the momentum operator $\boldsymbol{p}$, which is described by symmetry $\Gamma_4^-$, yields the characters of the optical transition driving operator $\boldsymbol{A} \cdot \boldsymbol{p}$. The resulting character sets are listed in Table II. The assignment of symmetries is done with the help of the character and multiplication tables for symmetry group $O_h$. Both tables as well as additional character sets for OAM light with different amount of orbital angular momentum $l$ as well as their symmetries are listed in the Supplemental Material in Table S2-S4.[24]

**TABLE II. Character set for OAM light field operator ($A = A_0 e^{il\varphi}$) and first-class (dipole, $A \cdot p = A_0 e^{il\varphi} \cdot p$) and second-class ($A \cdot p = A_0 \cdot ikz\, e^{il\varphi} \cdot p$) transition driving operator for OAM light with arbitrary amount of angular momentum *l*.**

| $O_h$ | E | $8C_3$ | $3C_2$ | $6C_4$ | $6C_2$' |
|---|---|---|---|---|---|
| $A = A_0 e^{il\varphi}$ | 6 | 0 | $2(-1)^l$ | $2\cos(l\pi/2)$ | 0 |
| $A \cdot p$ (first class) | 18 | 0 | $-2(-1)^l$ | $2\cos(l\pi/2)$ | 0 |
| $A \cdot p$ (second class) | 54 | 0 | $2(-1)^l$ | $2\cos(l\pi/2)$ | 0 |

Electric quadrupole transitions require a change of two units of angular momentum ($\Delta j = 2$) in matter and are sensitive to the light field gradient Q∇**E**. Usually, optical beams have a longitudinal field gradient, which allows for driving electric quadrupole transitions, however, with a strength of three orders of magnitude weaker than the electric dipole transition. A transverse gradient, due to the spatial structure of the beam front, such as in OAM light, can drive quadrupole transitions, too.[29-34] However, to make $\Delta j > 1$ transitions similar in magnitude to standard electric dipole transitions ($\Delta j = 1$), an atom usually has to be placed precisely in the vortex center (no further than an atomic size $a_0$) and the probe beam has to be focused close to the diffraction limit. In contrast, the micrometer length scales of Rydberg excitons in cuprite and focused OAM beams match, hence allowing for the realization of enhanced quadrupole transitions triggered by the transverse field gradient in the center of the OAM beam (see FIG. 1 (b)).[35] In order to predict whether such a second-class transition in excitons is allowed using OAM light, the symmetries of the optical transition driving operator are multiplied with the symmetry $\Gamma_4^-$, which accounts for the additional position vector in the quadrupole field. However, this way, we obtain the symmetry set for all second-class transitions, meaning including also magnetic-dipole transitions, resulting for $l = 0$ into: $\Gamma_1^+ + \Gamma_3^+ + \Gamma_4^+ + \Gamma_5^+$. From Ref. [24] we know that the only possible quadrupole transitions are of symmetry $\Gamma_3^+$ and $\Gamma_5^+$. The resulting character sets for first- and second-class transitions for an arbitrary amount of orbital angular momentum $l$ are listed in Table II. More detailed information can be found in Table S2 in the Supplemental Material.[24]

## B. Symmetry of excitons in cuprous oxide

The total symmetry of an exciton is determined by the product of valence band (hole), conduction band (electron) and envelope function (angular momentum $l^{exc}$) symmetries: $\Gamma_{exc} = \Gamma_h \times \Gamma_e \times \Gamma_{env}$. In cuprous oxide, the electronic configurations of the single atoms (Cu and O) determine that conduction and valence bands are mainly formed from the Cu 4s and Cu 3d functions, respectively.[25,36,37] In the crystal field the Cu 4s conduction band obtains symmetry $\Gamma_1^+$. The Cu 3d valence band splits into an energetically higher-lying, three-fold degenerate band of symmetry $\Gamma_5^+$ and a lower-lying, two-fold degenerate band of symmetry $\Gamma_3^+$. Due to the spin-orbit coupling between the quasispin $I$ and the hole spin $S_h$, the now six-fold (including spin) degenerate $\Gamma_5^+$ valence band splits into a higher lying two-fold degenerate band of symmetry $\Gamma_7^+$ and a lower lying four-fold degenerate band of symmetry $\Gamma_8^+$ by an amount of $\Delta = 130\ meV$.[38] Including spin, the $\Gamma_1^+$ conduction band is now two-fold degenerate and described by symmetry $\Gamma_6^+$. We consider here the yellow exciton series which occurs between the uppermost valence band ($\Gamma_7^+$) and the lowest conduction band ($\Gamma_6^+$) (see FIG. 1 (a)), thus $\Gamma_{exc} = \Gamma_h \times \Gamma_e \times \Gamma_{env} = \Gamma_7^+ \times \Gamma_6^+ \times \Gamma_{env}^{l^{exc}}$.

The symmetry property of the pure Coulomb field between electron and hole gives rise to the degeneracy of all levels with the same principal quantum number $n$ irrespective of their angular momentum quantum number $l^{exc}$, which is described by the exciton envelope function, expressed in spherical harmonic functions $Y_l^m$.[39] These are listed in Table S5 in the Supplemental Material in Cartesian coordinates for $l^{exc} = 0 \ldots 4$.[24] It suffices to use one representative function per orbital (symmetry group), i.e., choosing one magnetic quantum number value $m$, $Y_0^0$ for S or $Y_1^1$ for P, to perform symmetry considerations. In the simplest case, the orbital functions are pure basis functions, so their symmetry can be directly assigned according to the character table of symmetry group $O_h$. The orbital functions are then irreducible representations, which is the case for S- and P-orbitals: $Y_0^0 = 1/\sqrt{4\pi} \sim 1$ → S-orbital transforms as $\Gamma_1^+$; $Y_1^1 = -\sqrt{3/8\pi}\,(x - iy)/r$ → P-orbital transforms as $\Gamma_4^-$. If the bases cannot be seen directly, one has to apply the different symmetry operations of group $O_h$ to the orbital function and decompose them via $N_{m_1,m_2} = \langle Y_l^{m_1} | \hat{O} | Y_l^{m_2} \rangle$, yielding the complete character sets (shown in Table S6 in the Supplemental Material).[24] The resulting orbital symmetries, known via comparison of the character sets with the $O_h$ character table, are shown in Table III together with the complete exciton symmetries for different envelope function. In addition, the crystal ground state has symmetry $\Gamma_1^+$. Therefore the transition may be allowed if the symmetry of the excitonic state appears in the decomposition of the optical transition driving operator.

**TABLE III. Exciton envelope ($\Gamma_{env}^{l}$) and total exciton transition ($\Gamma_{exc}$ ) symmetries.**[28,40,41]

| Envelope $l^{exc}$ | Exciton envelope symmetry $\Gamma_{env}^{l^{exc}}$ | Exciton total symmetry $\Gamma_{exc}$ |
|---|---|---|
| S ($l^{exc} = 0$) | $\Gamma_1^+$ | $\Gamma_2^+ + \Gamma_5^+$ |
| P ($l^{exc} = 1$) | $\Gamma_4^-$ | $\Gamma_2^- + \Gamma_3^- + \Gamma_4^- + 2\Gamma_5^-$ |
| D ($l^{exc} = 2$) | $\Gamma_3^+ + \Gamma_5^+$ | $\Gamma_1^+ + 2\Gamma_3^+ + 3\Gamma_4^+ + \Gamma_5^+$ |
| F ($l^{exc} = 3$) | $\Gamma_2^- + \Gamma_4^- + \Gamma_5^-$ | $2\Gamma_1^- + \Gamma_2^- + 2\Gamma_3^- + 4\Gamma_4^- + 3\Gamma_5^-$ |
| G ($l^{exc} = 4$) | $\Gamma_1^+ + \Gamma_3^+ + \Gamma_4^+ + \Gamma_5^+$ | $\Gamma_1^+ + 2\Gamma_2^+ + 3\Gamma_3^+ + 4\Gamma_4^+ + 5\Gamma_5^+$ |

## C. Interaction of OAM light with Rydberg excitons: Modifying dipole selection rules

In Table IV the different light field operators for dipole and quadrupole OAM light are summarized. "$l = 1$"- and "$l = 3$"-OAM light exhibit the same symmetry. If the dipolar light field operator is of positive parity, the corresponding quadrupole light field operator is of negative parity and vice versa. For successive increase of OAM $l$, the parity changes alternatively. When illuminating cuprous oxide with dipolar light ($l = 0$), P-excitons are visible in the absorption spectrum. According to our calculations P-excitons can also become allowed transitions in quadrupole excitation using $l = 1$ or $l = 3$ OAM light. In contrast, the usually dipole-forbidden S-exciton states cannot only be driven by the quadrupole field of even OAM light ($l = 0, 2, 4$), but also by dipole transitions with odd OAM light ($l = 1, 3$). The same holds for D- and G-excitons, while F- and H-excitons follow the excitation rules of P-excitons. These results are summarized in Table V.

**TABLE IV. Symmetries of dipole and quadrupole transition driving operator of OAM light with different amount of orbital angular momentum $l$.**

| Orbital angular momentum $l$ | Dipole | Quadrupole |
|---|---|---|
| 0 (even) | $\Gamma_4^-$ | $\Gamma_3^+ + \Gamma_5^+$ |
| 1 (odd) | $\Gamma_1^+ + \Gamma_2^+ + 2\Gamma_3^+ + 2\Gamma_4^+ + 2\Gamma_5^+$ | $\Gamma_1^- + \Gamma_2^- + 2\Gamma_3^- + 4\Gamma_4^- + 4\Gamma_5^-$ |
| 2 (even) | $\Gamma_2^- + \Gamma_3^- + 2\Gamma_4^- + 3\Gamma_5^-$ | $2\Gamma_1^+ + \Gamma_2^+ + 3\Gamma_3^+ + 4\Gamma_4^+ + 3\Gamma_5^+$ |
| 3 (odd) | $\Gamma_1^+ + \Gamma_2^+ + 2\Gamma_3^+ + 2\Gamma_4^+ + 2\Gamma_5^+$ | $\Gamma_1^- + \Gamma_2^- + 2\Gamma_3^- + 4\Gamma_4^- + 4\Gamma_5^-$ |
| 4 (even) | $\Gamma_1^- + \Gamma_3^- + 3\Gamma_4^- + 2\Gamma_5^-$ | $\Gamma_1^+ + 2\Gamma_2^+ + 3\Gamma_3^+ + 3\Gamma_4^+ + 4\Gamma_5^+$ |

**TABLE V. Exciton transition symmetries and OAM light they can be excited with.**

| Envelope $l^{exc}$ | Exciton total symmetry $\Gamma_{exc}$ | Parity | $l^{dipole}$ | $l^{quadrupole}$ |
|---|---|---|---|---|
| S | $\Gamma_2^+ + \Gamma_5^+$ | + | odd | even |
| P | $\Gamma_2^- + \Gamma_3^- + \Gamma_4^- + 2\Gamma_5^-$ | - | even | odd |
| D | $\Gamma_1^+ + 2\Gamma_3^+ + 3\Gamma_4^+ + \Gamma_5^+$ | + | odd | even |
| F | $2\Gamma_1^- + \Gamma_2^- + 2\Gamma_3^- + 4\Gamma_4^- + 3\Gamma_5^-$ | - | even | odd |
| G | $\Gamma_1^+ + 2\Gamma_2^+ + 3\Gamma_3^+ + 4\Gamma_4^+ + 5\Gamma_5^+$ | + | odd | even |
| H | $\Gamma_1^- + 2\Gamma_2^- + 4\Gamma_3^- + 5\Gamma_4^- + 6\Gamma_5^-$ | - | even | odd |

## D. Exciton and Light Mode Spatial Extensions

We would like to predict for which principal quantum number $n$ a Rydberg exciton in cuprous oxide would most likely interact with focused OAM light, due to their largest spatial overlap. Therefore, the S-exciton envelope wavefunctions are visualized as a function of $n$ and their spatial overlap with a light mode of $l = 1$ orbital angular momentum is discussed. The excitonic radial wave function $r^2 R_{Cu_2O}^2(r)$ is calculated in analogy to the hydrogen radial wave function $R_{nl}$ and the light mode spatial extension is calculated based on the formula for Laguerre-Gauss modes for $l = 1$ at the focal point $z = 0$: $\left|u_{l=1}^{p=0}(r, \varphi = 0, z = 0)\right|^2$. The results are shown in FIG. 2 for S-exciton states with principal quantum number $n = 9 \ldots 14$. From here we see that S-exciton states with principal quantum number $n = 12$ show the largest overlap with the spatial mode of $l = 1$ OAM light when focused down to a spot of $w_0 = 250\, nm$ beam waist, which corresponds to a radius $r = 497\, nm$. See Supplemental Material for more details.[24]

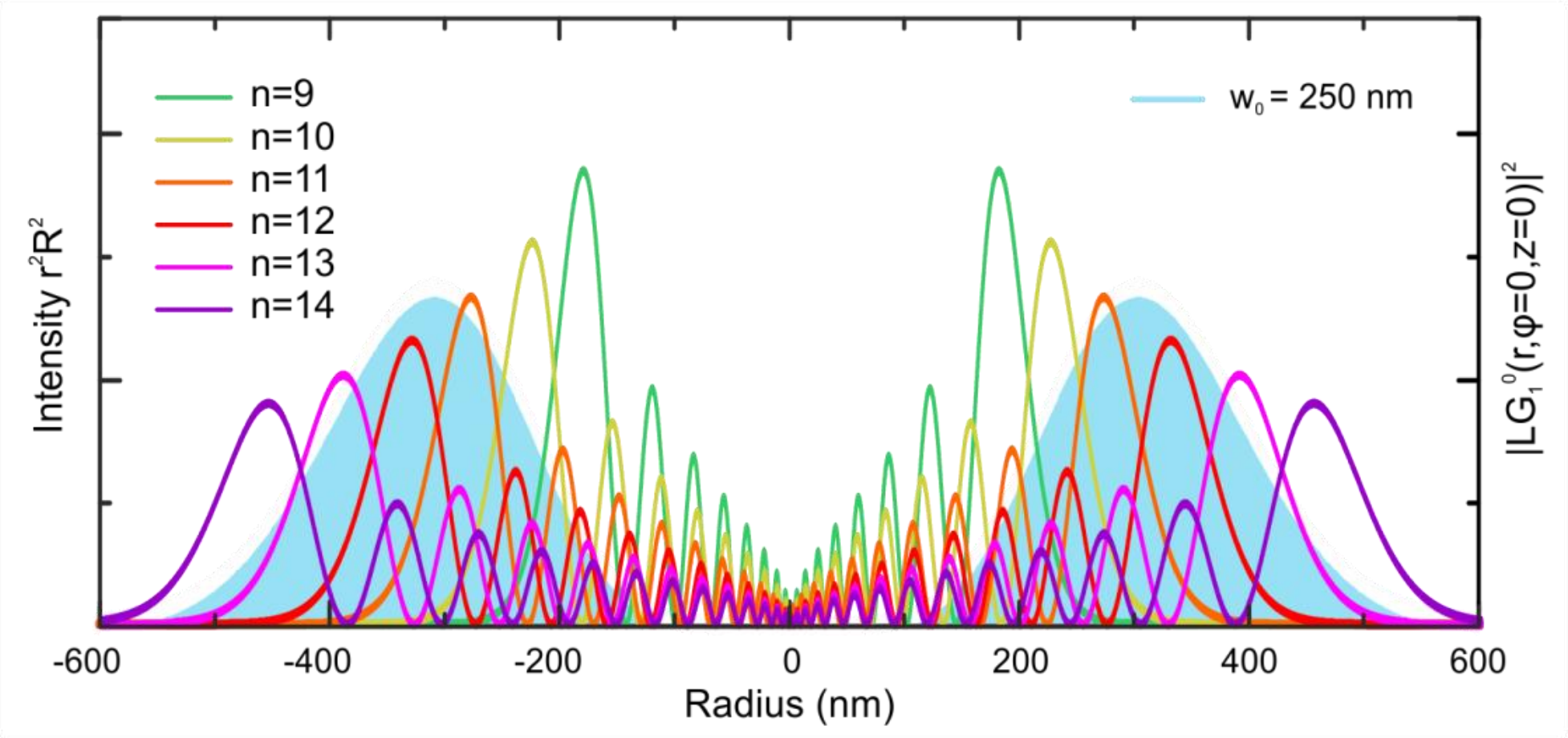

FIG. 2. Visualization of the spatial overlap between exciton states and a Laguerre-Gauss mode. The S-exciton ($l^{exc} = 0$) radial wave function $r^2 R^2_{Cu_2O}(r)$ is plotted for different quantum states $n = 9 \dots 14$. The intensity distribution of the Laguerre-Gauss mode ($l = 1$) is shown for a beam waist $w_0 = 250\ nm$.

## IV. CONCLUSION AND OUTLOOK

In the present paper we have given a detailed analysis of orbital angular momentum (OAM) light and exciton symmetries in cuprite. The symmetries of OAM light with different amount of OAM $l$ as well as the symmetries of excitons with different envelope functions (angular momentum quantum number $l^{exc}$) have been calculated. Comparing their overlap with the dipole selection rules allows us to predict that the alteration of the OAM of light may enable one to modify the optical transition selection rules and, hence, to excite usually dipole-forbidden Rydberg excitons in $Cu_2O$. We find that s- and d-envelope wavefunction excitons should be excitable with $l = 1$ and $l = 3$ OAM light. The precise oscillator strength of the transitions, however, requires further detailed theoretical investigation, meaning elaborate numerical calculations, taking the exact band structure, the exciton envelope wavefunction, and the exact spatial shape and extension of the OAM beam into account. However, it has been shown that the normally weak optical quadrupole interaction in *atoms* is enhanced significantly when the atom interacts at near resonance with an optical vortex.[32] Furthermore, transition amplitudes have been calculated for excitation of hydrogen-like atoms by OAM light.[31] If the target atom is located at distances of the order of atomic size near the phase singularity in the vortex center, the transition rates into states with OAM $l > 1$ become comparable with the rates for electric dipole transitions. As the Rydberg exciton is located in the center of the OAM beam and senses its complete light field, we assume the effect to be substantially enhanced in solids as well.

Using group-theoretical methods, we propose a new method for the observation of dipole-forbidden excitons in cuprite. In order to test the predictions we would like to implement the corresponding experiment as follows (see FIG. 1 (b)): Orbital angular momentum light can be created by imprinting a phaseplate on a fiber facet.[42] Upon transmission through the phaseplate, a beam of wavelength $\lambda$ is subjected to a phase delay $\psi$ which depends on the azimuthal angle $\varphi$, where $\psi = (n_{pp} - n_0)s\varphi/\lambda$ ($s$: step height, $n_0/n_{pp}$: refractive index of the surrounding material and phaseplate, respectively). A screw phase-dislocation produced on-axis causes destructive interference leading to the characteristic ring intensity pattern in the far field. For a pure Laguerre-Gauss mode the total phase delay around the phaseplate must be an integer multiple of $2\pi$, thus the physical step height in the spiral phaseplate is given by $s = l\lambda/(n_{pp} - n_0)$. The purity of Laguerre-Gaussian modes is limited by the co-production of higher order modes. This OAM light is then focused for example by a Fresnel lens carved into the $Cu_2O$ crystal, and transmission measurements are performed. We are going to analyze the different transition probabilities as set of quantum number $n$, exciton envelope wavefunction (angular momentum quantum number $l^{exc}$), as well as orbital angular momentum quantum number $l$. In addition, we want to calculate the transition matrix elements and oscillator strength using DFT. Then we would also be able to predict how probable an allowed transition will be.

Overall, only few experiments have been conducted so far with macroscopic Rydberg excitons in cuprite. Using OAM light will modify their absorption spectrum. In particular, D-excitons ($l = 2$) have not yet been investigated in detail in contrast to S- and P-excitons.[7,28,43,44] Furthermore, no experiments have been performed up to now, that combine OAM light and macroscopic quantum objects, such as the Rydberg excitons in $Cu_2O$.

## ACKNOWLEDGEMENTS

We gratefully acknowledge funding by the Deutsche Forschungsgemeinschaft (DFG) (SPP 1929 GiRyd) and the European Research Council (ERC) (Complexplas). Discussions with T. Pfau, M. M. Glazov, and J. Heckötter are acknowledged.

# Supplementary Information

## I. $O_H$ SYMMETRY GROUP

Cuprous oxide belongs to symmetry group $O_h$, which is of order 48 and contains all operations which transform a cube into itself. The 24 proper rotations are in the classes E: identity; $8C_3$: rotations of 2π/3 about the eight threefold space diagonal axes <111>, <-1-1-1>, <11-1>, <-1-11>, <1-11>, <-11-1>, <-111>, <1-1-1>; $3C_2$: rotations of π about the three cubic coordinate axes <100>, <010>, <001>; $6C_4$: rotations of π/2 about the cubic coordinate axes <100>, <-100>, <010>, <0-10>, <001>, <00-1>; $6C_2$': rotations of π about the six twofold face diagonal axes <110>, <1-10>, <101>, <10-1>, <011>, <01-1> (see FIG. S1). The 24 improper rotations are I: inversion; $8S_6$: rotations through π/3 about the eight space diagonal axes followed by a reflection in the plane perpendicular to the axis of rotation; $3\sigma_h$: reflection in a plane perpendicular to the principal axis of symmetry, i.e., xy-plane (ts0), xz-plane (t0r), yz-plane (0sr); $6S_4$: rotations through π/2 about the six coordinate axes followed by a reflection in the plane perpendicular to the axis of rotation; $6\sigma_d$: reflection containing a principal axis of symmetry that bisects the angle between two two-fold rotation axes perpendicular to the principal axis, i.e., ±45° to xy-, xz-, and yz-plane (tvv), (tv-v), (usu), (us-u), (wwr), (w-wr).[25]

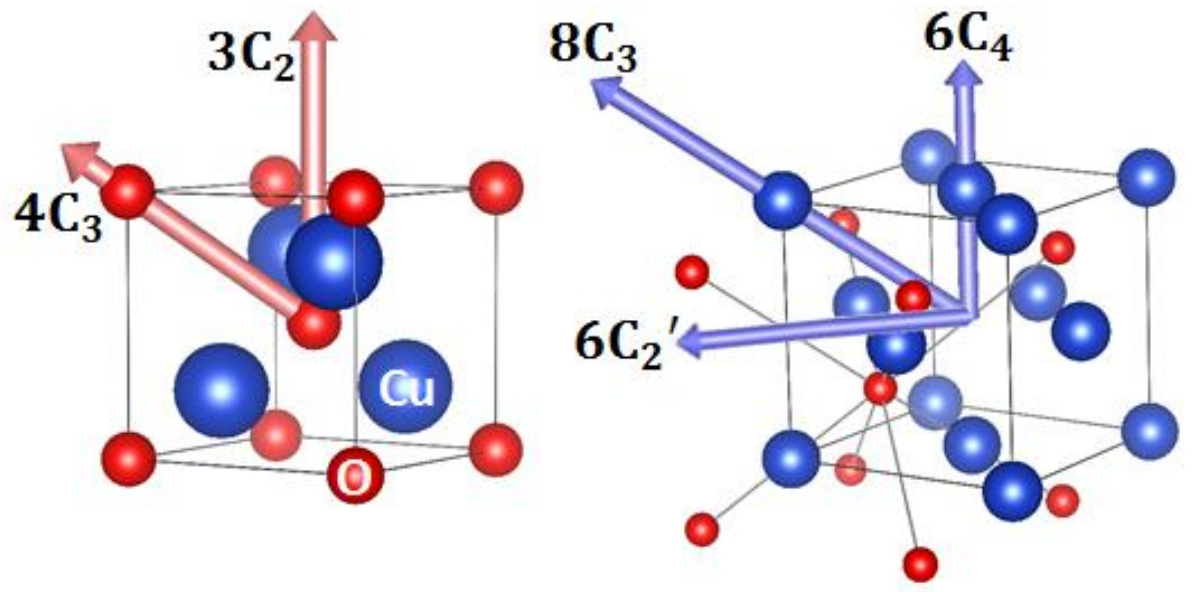

FIG. S1. $Cu_2O$ crystal lattice. O atoms (red) form a bcc lattice, Cu atoms (blue) form a fcc lattice. The rotational operations of the crystallographic groups T and O fit to the positions of the O- and Cu-atoms respectively. The number of similar operations are written in front of the rotation symbol.

## II. COORDINATE TRANSFORMATIONS UNDER $O_H$ SYMMETRY OPERATIONS

**TABLE S1. Coordinate transformations under $O_h$ symmetry operations.**

| Operator $\hat{O}$ | Vector | $x$ | $y$ | $z$ |
|---|---|---|---|---|
| E | | $x \to x$ | $y \to y$ | $z \to z$ |
| $8C_3$ | (111) | $x \to y \to z \to x$ | $y \to z \to x \to y$ | $z \to x \to y \to z$ |
| | (-1-1-1) | $x \to z \to y \to x$ | $y \to x \to z \to y$ | $z \to y \to x \to z$ |
| | (-111) | $x \to -z \to -y \to x$ | $y \to -x \to z \to y$ | $z \to y \to -x \to z$ |
| | (1-1-1) | $x \to -y \to -z \to x$ | $y \to z \to -x \to y$ | $z \to -x \to y \to z$ |
| | (1-11) | $x \to -y \to z \to x$ | $y \to -z \to -x \to y$ | $z \to x \to -y \to z$ |
| | (-11-1) | $x \to z \to -y \to x$ | $y \to -x \to -z \to y$ | $z \to -y \to x \to z$ |
| | (11-1) | $x \to -z \to -y \to x$ | $y \to x \to -z \to y$ | $z \to -y \to -x \to z$ |
| | (-1-11) | $x \to -y \to -z \to x$ | $y \to -z \to x \to y$ | $z \to -x \to -y \to z$ |
| $3C_2$ | (100) | $x \to x \to x$ | $y \to -y \to y$ | $z \to -z \to z$ |
| | (010) | $x \to -x \to x$ | $y \to y \to y$ | $z \to -z \to z$ |
| | (001) | $x \to -x \to x$ | $y \to -y \to y$ | $z \to z \to z$ |
| $6C_4$ | X | $x \to x \to x \to x \to x$ | $y \to -z \to -y \to z \to y$ | $z \to y \to -z \to -y \to z$ |
| | -X | $x \to x \to x \to x \to x$ | $y \to z \to -y \to -z \to y$ | $z \to -y \to -z \to y \to z$ |
| | Y | $x \to z \to -x \to -z \to x$ | $y \to y \to y \to y \to y$ | $z \to -x \to -z \to x \to z$ |
| | -Y | $x \to -z \to -x \to z \to x$ | $y \to y \to y \to y \to y$ | $z \to x \to -z \to -x \to z$ |
| | Z | $x \to -y \to -x \to y \to x$ | $y \to x \to -y \to -x \to y$ | $z \to z \to z \to z \to z$ |
| | -Z | $x \to y \to -x \to -y \to x$ | $y \to -x \to -y \to x \to y$ | $z \to z \to z \to z \to z$ |
| $6C_2$' | (110) | $x \to y \to x$ | $y \to x \to y$ | $z \to -z \to z$ |
| | (1-10) | $x \to -y \to x$ | $y \to -x \to y$ | $z \to -z \to z$ |
| | (101) | $x \to z \to x$ | $y \to -y \to y$ | $z \to x \to z$ |
| | (10-1) | $x \to -z \to x$ | $y \to -y \to y$ | $z \to -x \to z$ |
| | (011) | $x \to -x \to x$ | $y \to z \to y$ | $z \to y \to z$ |
| | (01-1) | $x \to -x \to x$ | $y \to -z \to y$ | $z \to -y \to z$ |

## III. TRANSFORMATION MATRIX DIAGONAL ELEMENTS AND RESULTING CHARACTERS

**TABLE S2. Matrix elements of the transformation matrices $\hat{O}|f_i\rangle = \sum_k N_{i,k}|f_k\rangle$ and their common characters for OAM light ($e^{il\varphi}$) with arbitrary amount of angular momentum *l*, as well as character sets for OAM light field operator $A = A_0 e^{il\varphi}$ and OAM first-class ($A \cdot p = A_0 e^{il\varphi} \cdot p$, dipole) and second-class ($A \cdot p = A_0 \cdot ikz\, e^{il\varphi} \cdot p$) transition driving operator.**

| $O_h$ | E | $8C_3$ | $3C_2$ | $6C_4$ | $6C_2$' | I | $8S_6$ | $3\sigma_h$ | $6S_4$ | $6\sigma_d$ |
|---|---|---|---|---|---|---|---|---|---|---|
| $e^{+il\varphi_x}$ | 1 | 0 | $e^{+il\pi}$ | $e^{+il\pi/2}$ | 0 | $e^{+il\pi}$ | 0 | 1 | $e^{+il\pi/2}$ | 0 |
| $e^{-il\varphi_x}$ | 1 | 0 | $e^{+il\pi}$ | $e^{+il\pi/2}$ | 0 | $e^{+il\pi}$ | 0 | 1 | $e^{+il\pi/2}$ | 0 |
| $e^{+il\varphi_y}$ | 1 | 0 | 0 | 0 | 0 | $e^{+il\pi}$ | 0 | 0 | 0 | 0 |
| $e^{-il\varphi_y}$ | 1 | 0 | 0 | 0 | 0 | $e^{+il\pi}$ | 0 | 0 | 0 | 0 |
| $e^{+il\varphi_z}$ | 1 | 0 | 0 | 0 | 0 | $e^{+il\pi}$ | 0 | 0 | 0 | 0 |
| $e^{-il\varphi_z}$ | 1 | 0 | 0 | 0 | 0 | $e^{+il\pi}$ | 0 | 0 | 0 | 0 |
| **$e^{il\varphi}$ (Sum)** | **6** | **0** | **$2(-1)^l$** | **$2\cos(l\pi/2)$** | **0** | **$6(-1)^l$** | **0** | **2** | **$2\cos(l\pi/2)$** | **0** |
| **$A_0$** | 1 | 1 | 1 | 1 | 1 | 1 | 1 | 1 | 1 | 1 |
| **$A = A_0 \cdot e^{il\varphi}$** | **6** | **0** | **$2(-1)^l$** | **$2\cos(l\pi/2)$** | **0** | **$6(-1)^l$** | **0** | **2** | **$2\cos(l\pi/2)$** | **0** |
| $l = 1$ | 6 | 0 | −2 | 0 | 0 | −6 | 0 | 2 | 0 | 0 |
| $l = 2$ | 6 | 0 | 2 | −2 | 0 | 6 | 0 | 2 | −2 | 0 |
| $l = 3$ | 6 | 0 | −2 | 0 | 0 | −6 | 0 | 2 | 0 | 0 |
| $l = 4$ | 6 | 0 | 2 | 2 | 0 | 6 | 0 | 2 | 2 | 0 |
| **$p$** | 3 | 0 | −1 | 1 | −1 | −3 | 0 | 1 | −1 | 1 |
| **$A \cdot p$ (first class)** | **18** | **0** | **$-2(-1)^l$** | **$2\cos(l\pi/2)$** | **0** | **$-18(-1)^l$** | **0** | **2** | **$-2\cos(l\pi/2)$** | **0** |
| $l = 1$ | 18 | 0 | 2 | 0 | 0 | 18 | 0 | 2 | 0 | 0 |
| $l = 2$ | 18 | 0 | −2 | −2 | 0 | −18 | 0 | 2 | 2 | 0 |
| $l = 3$ | 18 | 0 | 2 | 0 | 0 | 18 | 0 | 2 | 0 | 0 |
| $l = 4$ | 18 | 0 | −2 | 2 | 0 | −18 | 0 | 2 | −2 | 0 |
| **$z$** | 3 | 0 | −1 | 1 | −1 | −3 | 0 | 1 | −1 | 1 |
| **$A \cdot p$ (second class)** | **54** | **0** | **$2(-1)^l$** | **$2\cos(l\pi/2)$** | **0** | **$54(-1)^l$** | **0** | **2** | **$2\cos(l\pi/2)$** | **0** |
| $l = 1$ | 54 | 0 | −2 | 0 | 0 | −54 | 0 | 2 | 0 | 0 |
| $l = 2$ | 54 | 0 | 2 | −2 | 0 | 54 | 0 | 2 | −2 | 0 |
| $l = 3$ | 54 | 0 | −2 | 0 | 0 | −54 | 0 | 2 | 0 | 0 |
| $l = 4$ | 54 | 0 | 2 | 2 | 0 | 54 | 0 | 2 | 2 | 0 |

## IV. CHARACTER TABLE AND BASIS FUNCTIONS FOR THE GROUP $O_H$

**TABLE S3. Character table and basis functions for the group $O_h$.[23]**

| $O_h$ | E | $8C_3$ | $3C_2$ | $6C_4$ | $6C_2$' | I | $8S_6$ | $3\sigma_h$ | $6S_4$ | $6\ \sigma_d$ | Bases |
|---|---|---|---|---|---|---|---|---|---|---|---|
| $\Gamma_1^+$ | 1 | 1 | 1 | 1 | 1 | 1 | 1 | 1 | 1 | 1 | $R$ |
| $\Gamma_2^+$ | 1 | 1 | 1 | −1 | −1 | 1 | 1 | 1 | −1 | −1 | $(x^2-y^2)(y^2-z^2)(z^2-x^2)$ |
| $\Gamma_3^+$ | 2 | −1 | 2 | 0 | 0 | 2 | -1 | 2 | 0 | 0 | $(2z^2-x^2-y^2), \sqrt{3}(x^2-y^2)$ |
| $\Gamma_4^+$ | 3 | 0 | −1 | 1 | −1 | 3 | 0 | −1 | 1 | −1 | $S_x, S_y, S_z$ |
| $\Gamma_5^+$ | 3 | 0 | −1 | −1 | 1 | 3 | 0 | −1 | −1 | 1 | $yz, xz, xy$ |
| $\Gamma_1^-$ | 1 | 1 | 1 | 1 | 1 | −1 | −1 | −1 | −1 | −1 | $\Gamma_2^- \times \Gamma_2^+$ |
| $\Gamma_2^-$ | 1 | 1 | 1 | −1 | −1 | −1 | −1 | −1 | 1 | 1 | $xyz$ |
| $\Gamma_3^-$ | 2 | −1 | 2 | 0 | 0 | −2 | 1 | −2 | 0 | 0 | $\Gamma_3^+ \times \Gamma_2^-$ |
| $\Gamma_4^-$ | 3 | 0 | −1 | 1 | −1 | −3 | 0 | 1 | −1 | 1 | $x, y, z$ |
| $\Gamma_5^-$ | 3 | 0 | −1 | −1 | 1 | −3 | 0 | 1 | 1 | −1 | $\Gamma_5^+ \times \Gamma_1^-$ |
| $\Gamma_6^+$ | 2 | 1 | 0 | $\sqrt{2}$ | 0 | 2 | 1 | 0 | $\sqrt{2}$ | 0 | $\phi(1/2,-1/2), \phi(1/2,1/2)$ |
| $\Gamma_7^+$ | 2 | 1 | 0 | $-\sqrt{2}$ | 0 | 2 | 1 | 0 | $-\sqrt{2}$ | 0 | $\Gamma_6^+ \times \Gamma_2^+$ |
| $\Gamma_8^+$ | 4 | −1 | 0 | 0 | 0 | −4 | −1 | 0 | 0 | 0 | $\phi(3/2,-3/2), \phi(3/2,-1/2),$ $\phi(3/2,1/2), \phi(3/2,3/2)$ |
| $\Gamma_6^-$ | 2 | 1 | 0 | $\sqrt{2}$ | 0 | −2 | −1 | 0 | $-\sqrt{2}$ | 0 | $\Gamma_6^+ \times \Gamma_1^-$ |
| $\Gamma_7^-$ | 2 | 1 | 0 | $-\sqrt{2}$ | 0 | −2 | −1 | 0 | $\sqrt{2}$ | 0 | $\Gamma_6^+ \times \Gamma_2^-$ |
| $\Gamma_8^-$ | 4 | −1 | 0 | 0 | 0 | −4 | 1 | 0 | 0 | 0 | $\Gamma_8^+ \times \Gamma_1^-$ |

# V. MULTIPLICATION TABLE FOR THE GROUPS O AND $T_D$.

**TABLE S4. Multiplication table for the groups O and $T_d$. Results for group $O_h$ are obtained by taking into account parity, i.e., $\Gamma^{\pm} \times \Gamma^{\pm} = \Gamma^{+}$ and $\Gamma^{\pm} \times \Gamma^{\mp} = \Gamma^{-}$.[23]**

| $\Gamma_1$ | $\Gamma_2$ | $\Gamma_3$ | $\Gamma_4$ | $\Gamma_5$ | $\Gamma_6$ | $\Gamma_7$ | $\Gamma_8$ | $\times$ |
|---|---|---|---|---|---|---|---|---|
| $\Gamma_1$ | $\Gamma_2$ | $\Gamma_3$ | $\Gamma_4$ | $\Gamma_5$ | $\Gamma_6$ | $\Gamma_7$ | $\Gamma_8$ | $\Gamma_1$ |
| | $\Gamma_1$ | $\Gamma_3$ | $\Gamma_5$ | $\Gamma_4$ | $\Gamma_7$ | $\Gamma_6$ | $\Gamma_8$ | $\Gamma_2$ |
| | | $\Gamma_1+\Gamma_2+\Gamma_3$ | $\Gamma_4+\Gamma_5$ | $\Gamma_4+\Gamma_5$ | $\Gamma_8$ | $\Gamma_8$ | $\Gamma_6+\Gamma_7+\Gamma_8$ | $\Gamma_3$ |
| | | | $\Gamma_1+\Gamma_3+\Gamma_4+\Gamma_5$ | $\Gamma_2+\Gamma_3+\Gamma_4+\Gamma_5$ | $\Gamma_6+\Gamma_8$ | $\Gamma_7+\Gamma_8$ | $\Gamma_6+\Gamma_7+2\Gamma_8$ | $\Gamma_4$ |
| | | | | $\Gamma_1+\Gamma_3+\Gamma_4+\Gamma_5$ | $\Gamma_7+\Gamma_8$ | $\Gamma_6+\Gamma_8$ | $\Gamma_6+\Gamma_7+2\Gamma_8$ | $\Gamma_5$ |
| | | | | | $\Gamma_1+\Gamma_4$ | $\Gamma_2+\Gamma_5$ | $\Gamma_3+\Gamma_4+\Gamma_5$ | $\Gamma_6$ |
| | | | | | | $\Gamma_1+\Gamma_4$ | $\Gamma_3+\Gamma_4+\Gamma_5$ | $\Gamma_7$ |
| | | | | | | | $\Gamma_1+\Gamma_2+\Gamma_3+2\Gamma_4+2\Gamma_5$ | $\Gamma_8$ |

# VI. SPHERICAL HARMONICS AS EXCITON ENVELOPE FUNCTIONS

**TABLE S5. Spherical harmonics functions in Cartesian coordinates.**

| $l^{exc}$ | 0 | 1 | 2 | 3 | 4 |
|---|---|---|---|---|---|
| $m$ | S-exciton env. | P-exciton env. | D-exciton env. | F-exciton env. | G-exciton env. |
| $-4$ | | | | | $Y_4^{-4}(\theta,\varphi)=\frac{3}{16}\sqrt{\frac{35}{2\pi}}\frac{(x-iy)^4}{r^4}$ |
| $-3$ | | | | $Y_3^{-3}(\theta,\varphi)=\frac{1}{8}\sqrt{\frac{35}{\pi}}\frac{(x-iy)^3}{r^3}$ | $Y_4^{-3}(\theta,\varphi)=\frac{3}{8}\sqrt{\frac{35}{\pi}}\frac{(x-iy)^3 z}{r^4}$ |
| $-2$ | | | $Y_2^{-2}(\theta,\varphi)=\frac{1}{4}\sqrt{\frac{15}{2\pi}}\frac{(x-iy)^2}{r^2}$ | $Y_3^{-2}(\theta,\varphi)=\frac{1}{4}\sqrt{\frac{105}{2\pi}}\frac{(x-iy)^2 z}{r^3}$ | $Y_4^{-2}(\theta,\varphi)=\frac{3}{8}\sqrt{\frac{5}{2\pi}}\frac{(x-iy)^2(7z^2-r^2)}{r^4}$ |
| $-1$ | | $Y_1^{-1}(\theta,\varphi)=\frac{1}{2}\sqrt{\frac{3}{2\pi}}\frac{(x-iy)}{r}$ | $Y_2^{-1}(\theta,\varphi)=\frac{1}{2}\sqrt{\frac{15}{2\pi}}\frac{(x-iy)z}{r^2}$ | $Y_3^{-1}(\theta,\varphi)=\frac{1}{8}\sqrt{\frac{21}{\pi}}\frac{(x-iy)(4z^2-x^2-y^2)}{r^3}$ | $Y_4^{-1}(\theta,\varphi)=\frac{3}{8}\sqrt{\frac{5}{\pi}}\frac{(x-iy)z(7z^2-3r^2)}{r^4}$ |
| 0 | $Y_0^0(\theta,\varphi)=\frac{1}{2}\sqrt{\frac{1}{\pi}}$ | $Y_1^0(\theta,\varphi)=\frac{1}{2}\sqrt{\frac{3}{\pi}}\frac{z}{r}$ | $Y_2^0(\theta,\varphi)=\frac{1}{4}\sqrt{\frac{5}{\pi}}\frac{(2z^2-x^2-y^2)}{r^2}$ | $Y_3^0(\theta,\varphi)=\frac{1}{4}\sqrt{\frac{7}{\pi}}\frac{z(2z^2-3x^2-3y^2)}{r^2}$ | $Y_3^0(\theta,\varphi)=\frac{1}{4}\sqrt{\frac{7}{\pi}}\frac{z(2z^2-3x^2-3y^2)}{r^2}$ |
| 1 | | $Y_1^1(\theta,\varphi)=\frac{1}{2}\sqrt{\frac{3}{2\pi}}\frac{(x+iy)}{r}$ | $Y_2^1(\theta,\varphi)=-\frac{1}{2}\sqrt{\frac{15}{2\pi}}\frac{(x+iy)z}{r^2}$ | $Y_3^1(\theta,\varphi)=-\frac{1}{8}\sqrt{\frac{21}{\pi}}\frac{(x+iy)(4z^2-x^2-y^2)}{r^3}$ | $Y_4^1(\theta,\varphi)=-\frac{3}{8}\sqrt{\frac{5}{\pi}}\frac{(x+iy)z(7z^2-3r^2)}{r^4}$ |
| 2 | | | $Y_2^2(\theta,\varphi)=\frac{1}{4}\sqrt{\frac{15}{2\pi}}\frac{(x+iy)^2}{r^2}$ | $Y_3^2(\theta,\varphi)=\frac{1}{4}\sqrt{\frac{105}{2\pi}}\frac{(x+iy)^2 z}{r^3}$ | $Y_4^{-2}(\theta,\varphi)=\frac{3}{8}\sqrt{\frac{5}{2\pi}}\frac{(x+iy)^2(7z^2-r^2)}{r^4}$ |
| 3 | | | | $Y_3^3(\theta,\varphi)=-\frac{1}{8}\sqrt{\frac{35}{\pi}}\frac{(x+iy)^3}{r^3}$ | $Y_4^3(\theta,\varphi)=-\frac{3}{8}\sqrt{\frac{35}{\pi}}\frac{(x+iy)^3 z}{r^4}$ |
| 4 | | | | | $Y_4^4(\theta,\varphi)=\frac{3}{16}\sqrt{\frac{35}{2\pi}}\frac{(x+iy)^4}{r^4}$ |

# VII. EXCITON ENVELOPE FUNCTION CHARACTER SETS AND SYMMETRIES

**TABLE S6. Character sets and symmetries of exciton envelope functions.**

| **Envelope $l^{exc}$** | E (000) | $8C_3$ (111) | $3C_2$ (100) | $6C_4$ (X) | $6C_2$' (110) | **Exciton envelope symmetry $\Gamma_{env}^{l^{exc}}$** |
|---|---|---|---|---|---|---|
| S ($l^{exc}=0$) | 1 | 1 | 1 | 1 | 1 | $\Gamma_1^+$ |
| P ($l^{exc}=1$) | 3 | 0 | $-1$ | 1 | $-1$ | $\Gamma_4^-$ |
| D ($l^{exc}=2$) | 5 | $-1$ | 1 | $-1$ | 1 | $\Gamma_3^+ + \Gamma_5^+$ |
| F ($l^{exc}=3$) | 7 | 1 | $-1$ | $-1$ | $-1$ | $\Gamma_2^- + \Gamma_4^- + \Gamma_5^-$ |
| G ($l^{exc}=4$) | 9 | 0 | 1 | 1 | 1 | $\Gamma_1^+ + \Gamma_3^+ + \Gamma_4^+ + \Gamma_5^+$ |

# VIII. EXCITON AND LIGHT MODE SPATIAL EXTENSIONS

<u>Exciton radial wave function:</u>

$$R_{Cu_2O} = \frac{2}{n^2}\sqrt{\frac{(n-l-1)!}{(n+l)!^3}}\, e^{\frac{-\alpha_{Cu_2O} r}{2}} (\alpha_{Cu_2O} r)^l (-1)^{2l+1} \left(\frac{1}{\alpha_{Cu_2O}}\right)^{2l-1} \frac{\partial^{2l+1}}{\partial r^{2l+1}} \left( e^{\alpha_{Cu_2O} r} \left(\frac{1}{\alpha_{Cu_2O}}\right)^{n+l} \frac{\partial^{n+l}}{\partial r^{n+l}} \left( e^{-\alpha_{Cu_2O} r} (\alpha_{Cu_2O} r)^{n+l} \right) \right) \quad \text{(S1)}$$

$n$: principal quantum number; $l$: OAM quantum number; $\alpha_{Cu_2O} = \frac{2m^*/m_0}{\varepsilon a_0 n} = \frac{2\cdot 0.4}{9.8 a_0 n}$; $a_0 = \frac{4\pi\varepsilon_0\hbar^2}{m_0 e^2}$: Bohr radius.

<u>Laguerre-Gauss mode:</u>

$$u_l^p(r,\varphi,z) = \frac{C_{lp}^{LG}}{w(z)} \left(\frac{r\sqrt{2}}{w(z)}\right)^{|l|} e^{-\frac{r^2}{w^2(z)}} L_l^p\left(\frac{2r^2}{w^2(z)}\right) e^{-ik\frac{r^2}{2R(z)}} e^{il\varphi} e^{i\psi(z)} \quad \text{(S2)}$$

At the focus $z=0$: $R(z=0)=\infty$ and $e^{-ik\frac{r^2}{2R(z=0)}} = 1$, as well as $\psi(z=0) = \arctan(0) = 0$ and $e^{i\psi(z=0)} = 1$, and $w(z=0) = w_0$. Furthermore, we set $p=0$: $L_l^{p=0} = 1$ and $C_{lp}^{LG} = \sqrt{\frac{2p!}{\pi(p+|l|)!}} \to \sqrt{\frac{2}{\pi|l|!}}$. The Laguerre-Gauss mode for $l=1$ then becomes:

$$u_1^0(r,\varphi,z=0) = \frac{4r^3}{\sqrt{\pi} w_0^4} e^{-\frac{r^2}{w_0^2}} e^{i\varphi} \quad \text{(S3)}$$

As the mode is cylindrically symmetric, we set $\varphi = 0$ and plot the intensity distribution given by the absolute value squared as a function of $r$, $\left|u_{l=1}^{p=0}(r, \varphi = 0, z = 0)\right|^2$, for a beam waist $w_0 = 250\ nm$. The beam waist here refers to the thickness of the Laguerre-Gauss ring. We evaluate the maximum of the intensity cross section curve $\text{Max}\left|u_{l=1}^{p=0}(r, \varphi = 0, z = 0)\right|^2$ and calculate the radius at which the function is decayed to the $1/e^2$ of its maximum value, which we then define as the radius of the Laguerre-Gauss mode. For a beam waist $w_0 = 250\ nm$ this gives a radius $r = 497\ nm$.